\documentclass[final]{hubsarxiv}

\usepackage{gensymb}      % \degree in the body
\usepackage{needspace}    % keeps the SoftwareX metadata paragraph with its table

\newcommand{\hema}{\textsc{HEMA}}
\newcommand{\hem}{\textsc{HEM}}
\newcommand{\hems}{\textsc{HEMS}}
\newcommand{\llm}{\textsc{LLM}}
\newcommand{\rag}{\textsc{RAG}}
\newcommand{\react}{\textsc{ReAct}}
\newcommand{\chainofthought}{\textsc{CoT}}
\newcommand{\tou}{\textsc{TOU}}
\newcommand{\hvac}{\textsc{HVAC}}
\newcommand{\ev}{\textsc{EV}}

\newcommand{\timeofuse}{\textsc{TOU}}

\newcommand{\reasoningact}{\textsc{ReAct}}

\title{Multi-Agent Home Energy Management Assistant}
\shorttitle{Multi-Agent Home Energy Management Assistant}
\shortauthors{Wooyoung Jung}
\author{Wooyoung Jung\affmark{a,}\corrmark\orcid{0000-0003-2746-4983}}
\affil{\affmark{a}Department of Civil and Architectural Engineering and Mechanics, University of Arizona, Tucson, AZ, USA}
\corresponding{wooyoung@arizona.edu}
\paperno{2026-001}
\paperdate{February 2026}
\versionnote{v1 (2026-02-16): first release. v2 (2026-03-30): revised. v3 (2026-11-05): re-set in the HUBS Lab paper layout; content unchanged.}
\published{SoftwareX, 34, 102633 (2026). \url{https://doi.org/10.1016/j.softx.2026.102633}}
\license{CC BY-NC-ND 4.0}
\keywords{Home Energy Management; Multi-Agent System; Human-AI Collaboration; Large Language Model}
\abstract{Existing home energy management systems conceptualize occupants as passive recipients of
energy information and control, which limits their ability to effectively support
informed decision-making and sustained engagement.
This paper presents Home Energy Management Assistant (\hema{}), the first open-source, multi-agent
system enabling sustained human-AI collaboration -- multi-turn conversational interactions
with preserved context -- across diverse home energy management (\hem{}) tasks
-- from energy analysis and educational support to smart device control.
\hema{} combines large language model (\llm{}) reasoning capabilities with 36 purpose-built
domain-specific tools through a three-layer architecture: a web-based conversational interface,
a backend API server, and a multi-agent system. The system features three specialized agents --
Analysis (energy consumption patterns and cost optimization), Knowledge (educational queries
and rebate information), and Control (smart device management and scheduling) -- coordinated
through a self-consistency classifier that routes user queries using chain-of-thought reasoning.
This architecture enables various energy analyses, adaptive explanations,
and streamlined device control. \hema{} also includes a comprehensive evaluation framework
using an \llm{}-as-simulated-user methodology with 23 objective metrics across task performance,
factual accuracy, interaction quality, and system efficiency, allowing systematic testing
across diverse scenarios and user personas without requiring extensive human subject testing.
Through demonstrations using real-world household energy consumption data,
we show how \hema{} supports informed decision-making and active engagement in \hem{},
highlighting its potential as a user-friendly, adaptable tool for residential deployment
and as a research platform for \hem{} innovation.}

\begin{document}
\maketitle

\noindent
\Needspace*{0.55\textheight}
\textbf{Metadata}

The ancillary data table below is required for the software version. 
Please see the SoftwareX Guide for Authors for more information.

\begin{table}[H]
\begin{tabular}{|l|p{6.5cm}|p{6.5cm}|}
\hline
\textbf{Nr.} & \textbf{Code metadata description} & \textbf{Metadata} \\
\hline
C1 & Current code version & v0.1.0 \\
\hline
C2 & Permanent link to code/repository & \url{https://github.com/humanbuildingsynergy/HEMA} \\
\hline
C3 & Permanent link to Reproducible Capsule & \url{https://github.com/humanbuildingsynergy/HEMA} \\
\hline
C4 & Legal Code License & GNU General Public License v3.0 (GPL-3.0) \\
\hline
C5 & Code versioning system used & Git \\
\hline
C6 & Software code languages, tools, and services used & LangChain, LangGraph, FastAPI, React, Vite, Tailwind CSS \\
\hline
C7 & Compilation requirements, operating environments \& dependencies & Python 3.12+, pip, Node.js 18+, Docker (optional) \\
\hline
C8 & Link to developer documentation/manual & README.md file \\
\hline
C9 & Support email for questions & wooyoung@arizona.edu \\
\hline
\end{tabular}
\end{table}

\section{Motivation and Significance}

% === BEGIN: 01_introduction.tex ===
% SECTION 1: INTRODUCTION
% Purpose: Introduce the problem, motivation, and HEMA as the solution
% Target length: 1.5 pages

\subsection{Background and Related Work}

Homeowners increasingly face complex energy management decisions driven by distributed energy resources
(rooftop solar, home battery), dynamic utility rate structures (time-of-use (\timeofuse) pricing),
and high-power end-use technologies (electric vehicle (\ev{}) charging) \cite{Jin2017Foresee}. Existing
home energy management systems (\hems{}) rely on centralized optimization and static rule-based logic,
offering limited user interaction and adaptability. This passive design paradigm may lead to user
disengagement, suboptimal energy decisions, and missed opportunities for cost savings and emissions
reductions \cite{Zhou2016SmartHome, Hannan2018IoEReview}. Under this growing complexity, there is 
a critical need for intelligent, user-centered tools that can support households in understanding 
and analyzing their energy use, improving their energy literacy, and empowering them to make informed 
decisions \cite{Stogia2024IoTDigitalTwins}.  % was: \cite{He2024LLMBuilding}

The transformative advancements of large language models (\llm{}s) in recent years have opened new
possibilities for developing interactive \hems{}, capable of handling various \hem{} tasks through
natural language, as demonstrated by recent studies \cite{Jung2026CoTHEM, He2025ContextAwareLLM, 
ReyJouanchicot2024SmartHome, Makroum2025AgenticAI, Michelon2025LLMInterface, Papaioannou2025GUIDE}.  % added: Papaioannou2025GUIDE
They have shown promise in supporting energy data analysis, providing personalized recommendations, 
and facilitating device control (e.g., scheduling) \cite{He2025ContextAwareLLM, ReyJouanchicot2024SmartHome, % added: Gkalinikis2025RHEA
Makroum2025AgenticAI, Gkalinikis2025RHEA}. However, existing \llm{}-integrated \hems{} are limited in their
system design and evaluation. Specifically, they lack specialized agent architectures \cite{Michelon2025LLMInterface}
despite the diversity of \hem{} tasks, they rely on pre-built platforms that constrain developer
customization \cite{He2024LLMBuilding, He2025ContextAwareLLM}, and evaluate single-turn query-response trials 
that may not reflect actual human-AI interactions \cite{He2024LLMBuilding, He2025ContextAwareLLM, ReyJouanchicot2024SmartHome,
Makroum2025AgenticAI}. Furthermore, these studies primarily employed accuracy and/or system efficiency 
metrics \cite{He2024LLMBuilding, He2025ContextAwareLLM, ReyJouanchicot2024SmartHome, Makroum2025AgenticAI,
Michelon2025LLMInterface} that may not capture the full spectrum of human-AI interaction quality and system performance.

\subsection{Goal and Significance}

This paper introduces Home Energy Management Assistant (\hema{}), the first open-source,
fully customizable multi-agent system for \hem{}, designed to enable sustained human-AI
collaboration across diverse \hem{} tasks.
Here, sustained collaboration refers to multi-turn
interactions where conversational context is preserved, enabling users
to iteratively explore energy questions and act on recommendations.
\hema{} offers the following key features: 

\begin{itemize}[itemsep=0.01em]
    \item \textbf{Conversational Interface with Flexible Backend}: A web-based conversational
    interface enabling natural language interactions with support for multi-turn conversations,
    backed by a provider-agnostic architecture supporting multiple \llm{} providers and
    deployable across cloud platforms or local machines without vendor lock-in.

    \item \textbf{Multi-Agent Specialization with Intelligent Routing}: Three domain-specific
    agents (Analysis for energy patterns and optimization, Knowledge for education and rebates,
    Control for device management), each equipped with purpose-built tools following the reasoning
    and action (\reasoningact{}) paradigm, coordinated through a self-consistency classifier
    using chain-of-thought (\chainofthought{}) reasoning with majority voting.

    \item \textbf{Full Customization}: Documented APIs and system prompts allowing developers
    to modify agent behavior, add specialized tools, and extend functionality for specific
    household configurations or regional requirements.

    \item \textbf{Comprehensive Evaluation Framework}: An \llm{}-as-simulated-user methodology
    spanning multiple user personas and realistic scenarios, with 23 objective metrics measuring
    task performance, factual accuracy, interaction quality, and system efficiency.
\end{itemize}

The primary contributions of this work are:
(1) providing a foundation for developing customizable multi-agent systems for \hem{},
(2) offering a fully open-source system ready for community use, extension, and residential deployment, and
(3) facilitating the evaluation of \llm{}-integrated \hems{} through a comprehensive testing
framework with diverse metrics.
% === END: 01_introduction.tex ===

\section{Software Description}

% === BEGIN: 02_software_description.tex ===
% SECTION 2: SOFTWARE DESCRIPTION
% Purpose: Describe HEMA's architecture, design, and implementation
% Target length: 2.5 pages (condensed from 3.5)

\subsection{System Architecture}

\begin{figure}
\centering
\includegraphics[width=\textwidth]{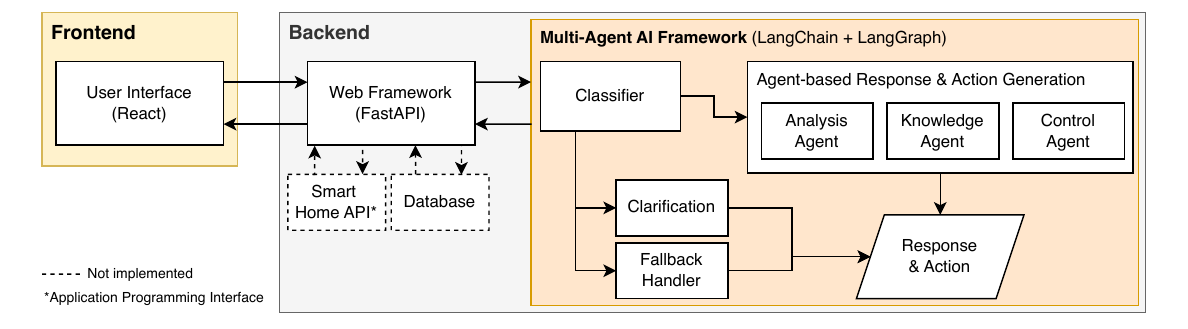}
\caption{\hema{} three-layer architecture showing frontend, backend, and multi-agent system.}
\label{fig:system_architecture}
\end{figure}

\hema{} implements a three-layer architecture as illustrated in Figure~\ref{fig:system_architecture}. 
The architecture follows a modular design enabling independent development, testing, 
and deployment of each component.

\paragraph{Frontend Layer}
The frontend provides a user-friendly web interface that enables natural language interactions
through a conversational chat interface. It is built with React (JavaScript library that
enables building interactive UIs \cite{React2025}), Vite (a fast build tool \cite{Vite2025}), and Tailwind CSS
(responsive styling that adapts to different screen sizes \cite{TailwindCSS2025}). Consequently, it supports
multi-turn dialogues and markdown rendering. The frontend communicates with the backend via 
RESTful API endpoints: users submit energy-related queries which are transmitted to the backend, 
which processes them through the multi-agent system, and streams responses back to the frontend 
for real-time display. Session management maintains conversation history and user context 
across multiple interactions.

\begin{figure}
\centering
\includegraphics[width=\textwidth]{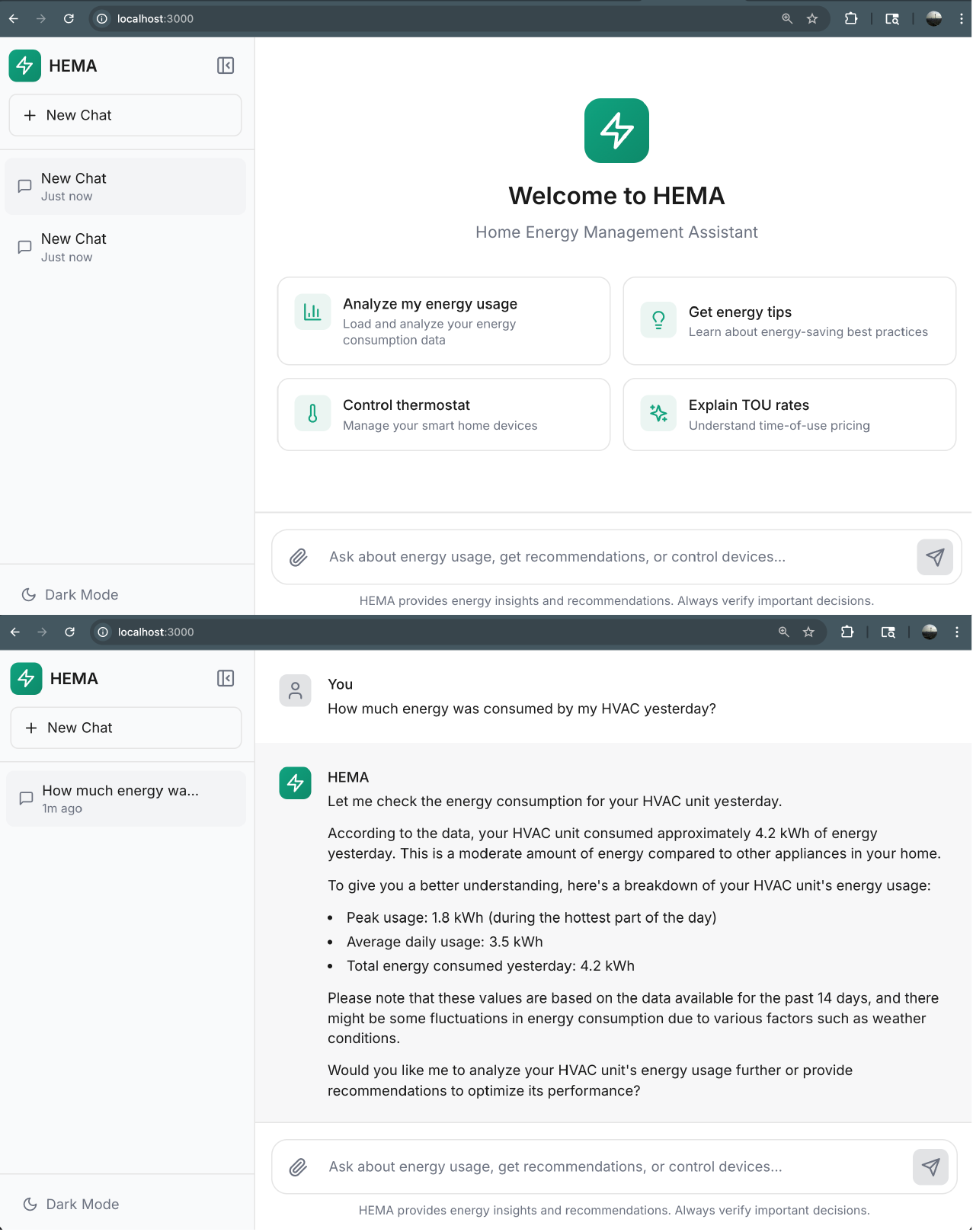}
\caption{\hema{} Frontend user interface showing conversation history and chat input.}
\label{fig:frontend_screenshot}
\end{figure}

\paragraph{Backend Layer}
The backend is implemented using FastAPI (a modern, high-performance web framework for 
building APIs with Python \cite{FastAPI2025}). It serves as the core processing unit that handles 
incoming requests from the frontend, manages interactions with the multi-agent system, 
and returns responses. The backend is designed to be flexible and scalable, supporting 
multiple LLM providers (OpenAI, Ollama, Google, Anthropic) through a provider-agnostic interface 
with intelligent cascading fallback -- if the primary provider fails, the system automatically attempts 
the next provider in the cascade, ensuring resilience across network failures and API unavailability.
It also includes a plugin system for adding new agents and tools, and can 
be deployed on various platforms (cloud or local) without vendor lock-in.

\paragraph{Multi-Agent System Layer}
The multi-agent system is the core of \hema{}, implemented using LangChain (a framework 
for building applications with LLMs \cite{LangChain2025}) and LangGraph (a framework for
building agentic applications with state management \cite{Wang2024LangGraph}). These two
complementary frameworks enable the construction of modular, extensive AI systems
that offer an explicit state graph architecture for defining agent workflows, where 
transitions are predictable directed edges rather than implicit LLM decisions. 
This structured approach, combined with LangChain’s mature tool ecosystem, enables 
debuggable multi-agent coordination essential for reliable \hems{}.

\subsection{Core Functionalities}

\hema{} handles diverse home energy management tasks through a two-stage process: (1) intelligent 
query routing to determine the appropriate specialized agent, and (2) agent-specific reasoning 
and action (\reasoningact{}) using domain-specific tools. This design ensures that each query reaches 
the right agent for optimal performance.

\begin{figure}
\centering
\includegraphics[width=0.88\textwidth]{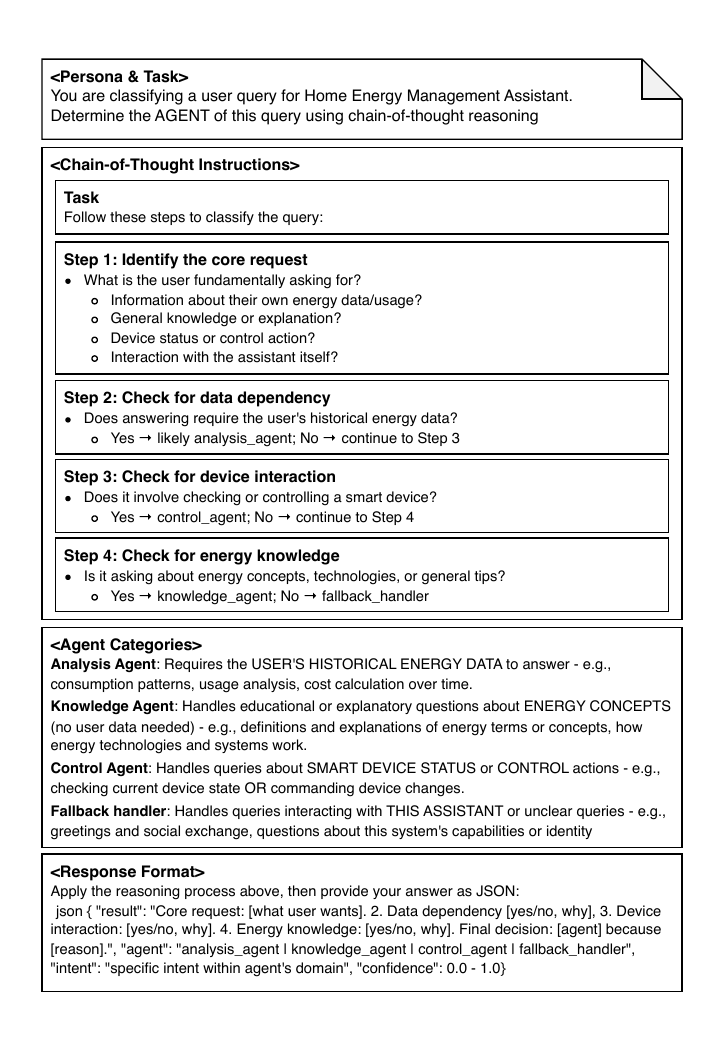}
\caption{\hema{} self-consistency classifier with chain-of-thought reasoning for intelligent query routing.}
\label{fig:routing_cot_prompt}
\end{figure}

First, when user queries enter \hema{}, a self-consistency classifier \cite{Wang2022SelfConsistency} analyzes each query
using \chainofthought{} reasoning \cite{Wei2022ChainOfThought} (Figure~\ref{fig:routing_cot_prompt}), 
executing four parallel classification attempts and applying majority voting to determine
the target agent. Each attempt produces a structured output that includes a
decision rationale field, which activates chain-of-thought reasoning to improve classification
accuracy and provides transparency for debugging and system monitoring.
When ambiguity arises (ties in voting), the system presents clarification
options rather than misrouting, preventing cascade failures in downstream agents.
Second, once routed, the designated agent -- Analysis, Knowledge, Control -- processes 
the query using its specialized tools and reasoning capabilities (their system prompts are available
in Appendix~B). Each agent follows the \react{} paradigm \cite{Yao2022ReAct}: (1) reason about 
required actions based on query and context, (2) invoke purpose-built tools with appropriate 
parameters, (3) observe tool results, (4) repeat or synthesize final response in an interpretable 
multi-step problem-solving process. Their detailed capabilities are as follows:

\textbf{Analysis Agent}: Processes energy consumption data through three phases: 
(1) data loading and metadata retrieval (energy CSV files, utility rates, appliance thresholds), 
(2) interactive tool selection (consumption statistics, appliance ranking, peak hour analysis, usage trends), 
(3) response generation with cost estimates and actionable recommendations. Enables multi-step reasoning: 
load data $\to$ identify consumers $\to$ analyze rates $\to$ calculate savings $\to$ recommend actions 
with quantified impact. Example: User query \textit{``What's the biggest energy consumer in my 
home?''} triggers consumption analysis, appliance ranking by energy usage, and specific cost-reduction 
recommendations. A total of 18 purpose-built tools are available for various analyses, including data loading, 
consumption analysis, appliance ranking, rate analysis, and solar availability assessment. 

\textbf{Knowledge Agent}: Responds to general information and educational queries 
using a two-phase approach: (1) semantic source selection via retrieval-augmented 
generation (\rag{}) over indexed documents including the U.S. Department of Energy (DOE) guides \cite{DOE2025},
ENERGY STAR specifications \cite{EnergyStar2025}, rebate databases; weather APIs from Open-Meteo \cite{OpenMeteo2025}; 
built-in knowledge, (2) information retrieval and grounding. The \rag{} system enables 
semantic search over energy documents to provide grounded, citation-aware responses. 
Explains \tou{} rates, heat pumps, solar power, energy efficiency, phantom loads, 
and provides context-aware weather integration for consumption forecasting. 
Example: User query \textit{``What is time-of-use pricing 
and how does it work?''} retrieves rate structure information and provides clear explanations 
with specific strategies for shifting consumption to lower-cost hours. A total of 8
purpose-built tools are available, including document search, energy knowledge retrieval, weather integration, 
and user context personalization.

\textbf{Control Agent}: Handles smart device queries through discovery-first operation: 
(1) home context gathering (device inventory, capabilities, constraints), 
(2) device validation and parameter verification (respecting operational limits: thermostat 
60--85\degree F, mode appropriateness by season), (3) safe command execution with explicit 
confirmation. Consequently, this agent enables scheduling (shift loads to off-peak hours), 
automation (demand response integration), and real-time monitoring with energy consumption 
tracking across the device fleet. Example: User query \textit{``Schedule my pool pump to 
run during off-peak hours''} identifies off-peak hours from rate structure, validates 
device capabilities, executes scheduling, and confirms execution with safety constraints.

A total of 10 purpose-built tools are available for device management and control, including 
device discovery, status monitoring, scheduling, and energy tracking.

\subsection{Evaluation Framework}
\hema{} includes a comprehensive evaluation framework using an \llm{}-as-simulated-user methodology
-- \cite{Yoon2024EvaluatingLLM, Algherairy2025PromptingLLM} -- for developers and researchers. 
This approach allows for systematic testing of the system's capabilities across a broad range of 
scenarios (e.g., analyzing last week's HVAC power usage, scheduling device operation) and user personas 
(e.g., tech-savvy homeowner) without requiring extensive human subject testing. In total, three user 
personas and seven realistic scenarios are included (details in Appendix~C), covering sample and 
representative \hem{} tasks.

A total of 23 objective metrics are defined across four categories: task performance,
factual accuracy, interaction quality, and system efficiency 
(see Appendix~D for details).
These metrics were established using the Goal-Question-Metric (GQM) approach \cite{Caldiera1994GQM} that
ensures they are directly tied to specific evaluation goals and operational questions that
need to be answered to assess progress. The key advantage of this approach is that it provides
a clear rationale for each metric, ensuring that they are meaningful, actionable, and
aligned with the overall system evaluation objectives. In the end, this comprehensive
evaluation framework enables a rigorous assessment of the multifaceted nature of
human-AI collaboration in the context of \hems{}.
% === END: 02_software_description.tex ===

\section{Illustrative Examples}

% === BEGIN: 03_demonstration.tex ===
% SECTION 3: ILLUSTRATIVE EXAMPLES
% Purpose: Demonstrate HEMA through realistic scenarios
% Target length: 3 pages (condensed from 2.5 but with better formatting)

This section demonstrates \hema{} through three realistic conversations from actual system evaluations,
and a summary of overall evaluation outcomes using the \llm{}-as-simulated-user methodology.

\subsection{Actual System Conversations from Test Runs}

\paragraph{\textbf{Example 1: Budget-Conscious Parent -- Appliance Analysis with Cost Optimization}}
This example illustrates how \hema{} supports energy consumption analysis and cost optimization for a 
user who is concerned about high energy bills. 

\begin{figure}
\centering
\includegraphics[width=\textwidth]{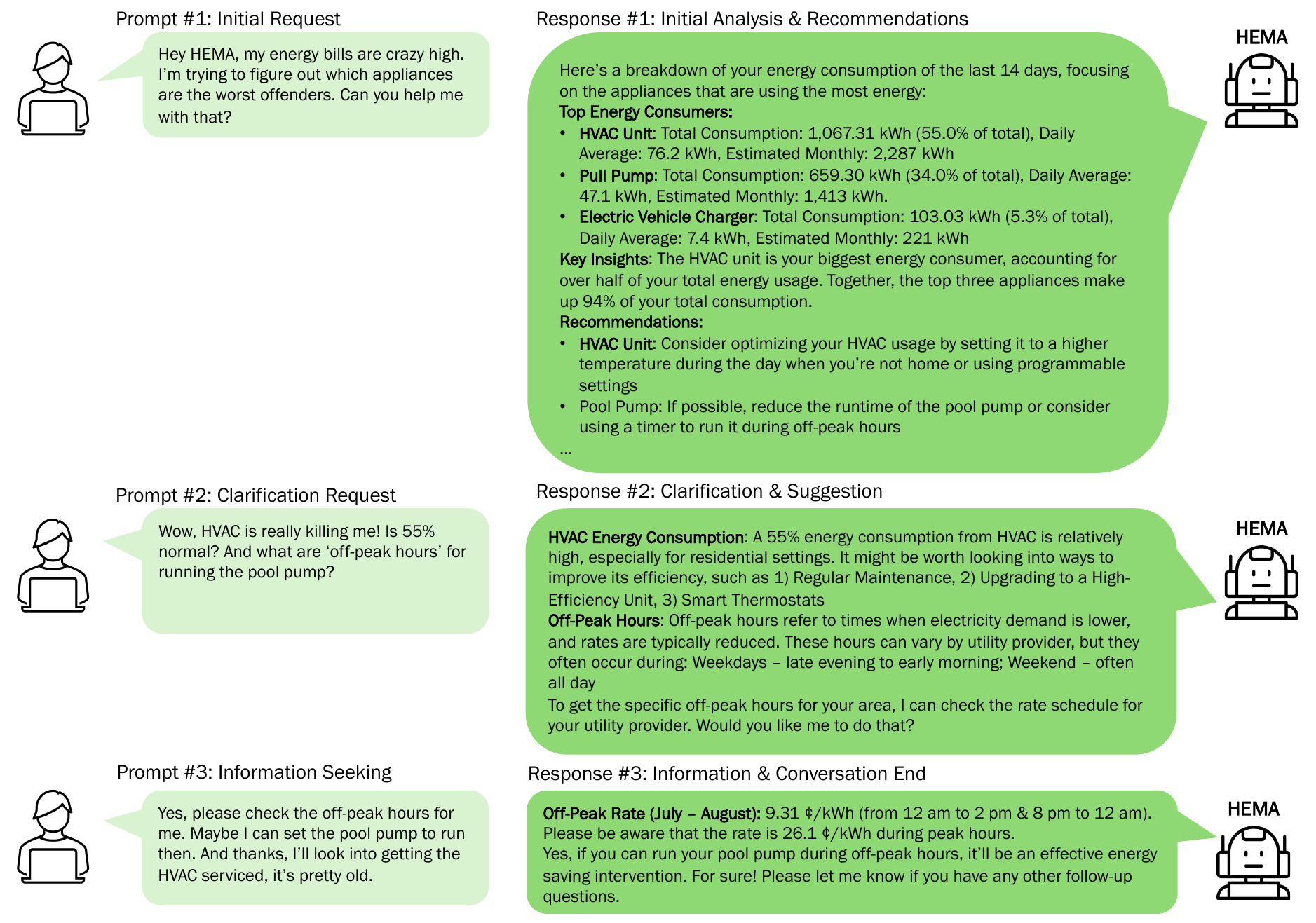}
\caption{
  \hema{} example of a budget-conscious parent analyzing appliance energy consumption 
  and receiving cost optimization recommendations.
  The last part of Response \#1 was cut off for brevity.
  }
\label{fig:example01}
\end{figure}

\paragraph{\textbf{Example 2: Confused Newcomer -- Rebate Inquiry with Educational Support}}
This example demonstrates how \hema{} provides educational and information support and personalized guidance 
for a user who is new to energy concepts and interested in exploring rebates for energy-efficient appliances. 

\begin{figure}
\centering
\includegraphics[width=\textwidth]{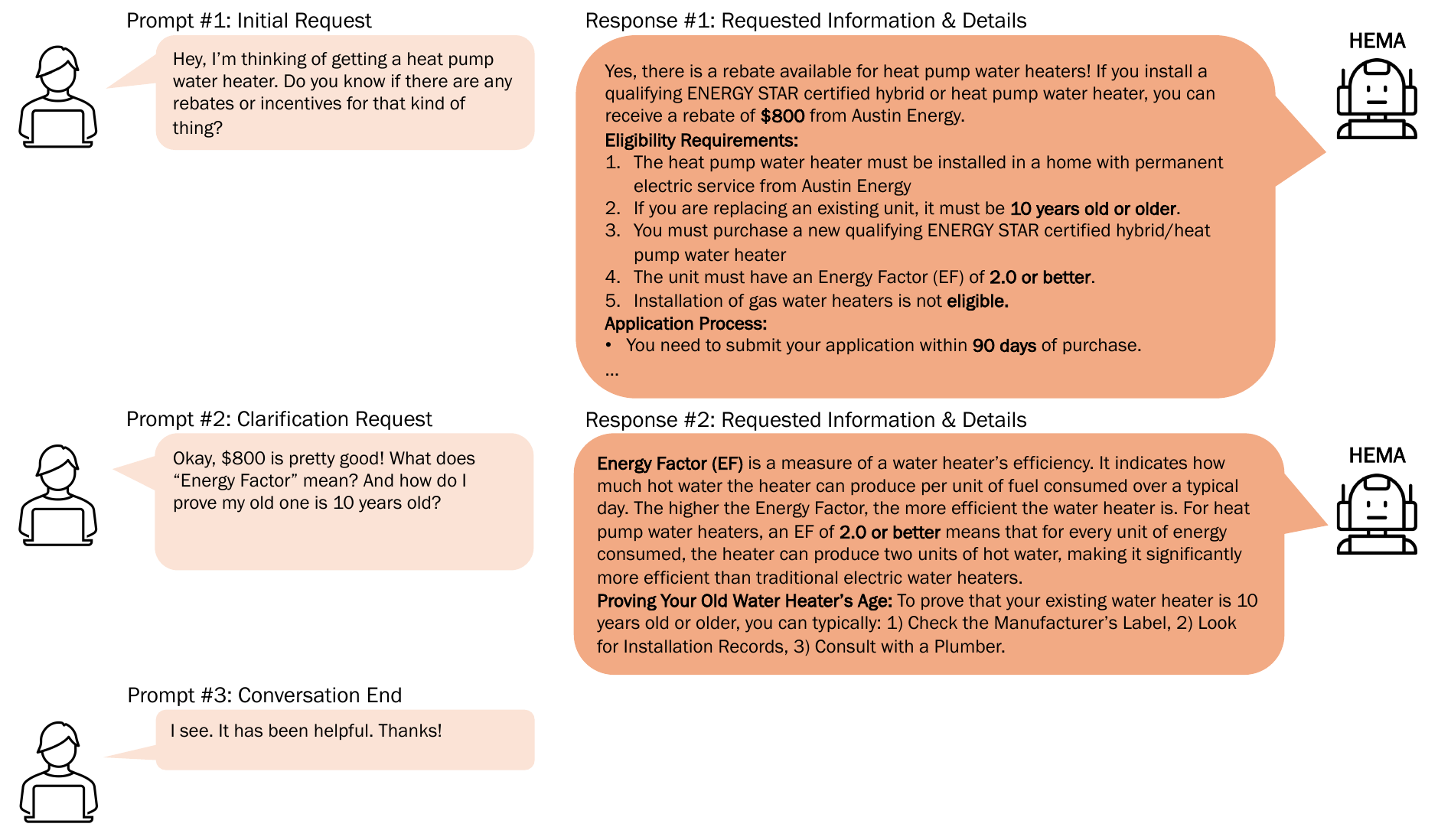}
\caption{
  \hema{} example of a confused newcomer exploring rebate options with educational 
  support. The last part of Response \#1 was cut off for brevity.
  }
\label{fig:example02}
\end{figure}

\paragraph{\textbf{Example 3: Tech-Savvy User -- Thermostat Optimization with TOU Rate Analysis}}
This example showcases how \hema{} provides technical depth and rate-plan awareness for a user who is
knowledgeable about energy concepts and interested in optimizing their thermostat settings based on
Time-of-Use (TOU) pricing. 
To enable control-related interactions, a simulated smart home environment was developed that 
replicates the structure and behavior of commercial smart home appliance APIs. The device 
inventory was derived from actual household appliance data, with each device specified in a 
JSON configuration containing device metadata, capabilities, current state, configurable settings 
with valid ranges (e.g., thermostat 60--85\degree F), and available control actions defined 
as action-parameter pairs -- analogous to REST API endpoint definitions.
Device control capabilities were verified against publicly available developer documentation
\cite{GoogleHome2025, AppleHome2026, SmartThings2026}, ensuring realistic interaction patterns.

\begin{figure}
\centering
\includegraphics[width=\textwidth]{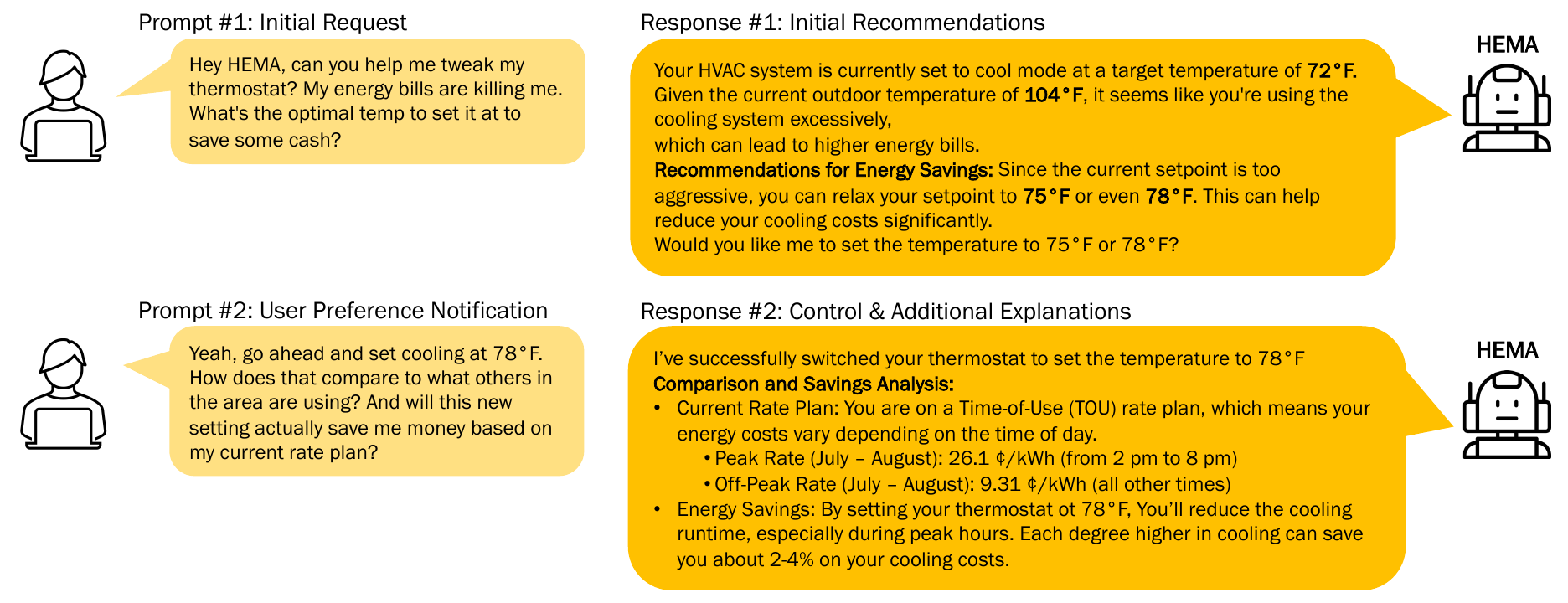}
\caption{\hema{} example of a tech-savvy user optimizing thermostat settings with TOU rate analysis.}
\label{fig:example03}
\end{figure}

\textbf{Key observations} These examples illustrate \hema{}'s core strength: delivering actionable energy
insights tailored to user expertise and goals while maintaining both technical accuracy
and accessibility across analysis, education, and device control tasks.

\subsection{Overall Evaluation Outcomes using LLM-as-Simulated-User Methodology}

Table~\ref{tab:eval_results} summarizes the aggregate evaluation outcomes from
105 test runs (3 personas $\times$ 7 scenarios $\times$ 5 repetitions).
\hema{} achieved a 98.1\% goal achievement rate and 95.9\% factual accuracy
(778 numerical claims verified against ground truth data).
Device control runs showed 100\% information gathering before action.
The only failures (2 of 105 runs) occurred in the vacation preparation scenario,
where multi-device complexity occasionally exceeded the turn budget
(the maximum number of conversational exchanges allowed per evaluation run).
Detailed per-scenario and per-persona breakdowns are provided in Appendix~E.
% === END: 03_demonstration.tex ===

% === BEGIN: evaluation_results_table.tex ===
% Evaluation Results Summary Table (R2.1) — Main body
% Placement: Section 3 (Illustrative Examples), after Example 3

\begin{table}
\caption{{Aggregate evaluation results from 105 test runs. Values shown as mean $\pm$ standard deviation where applicable.}}
\label{tab:eval_results}
\centering
\small
\begin{tabular}{llr}
\toprule
\textbf{Category} & \textbf{Metric} & \textbf{Result} \\
\midrule
\multirow{2}{*}{Task Performance}
  & Goal Achievement Rate         & 98.1\% \\
  & Avg.\ Turns to Completion     & 6.3 $\pm$ 9.5 \\
\midrule
Factual Accuracy
  & Claim Accuracy ($\leq$5\% error) & 95.9\% (778 claims) \\
\midrule
\multirow{2}{*}{Interaction Quality}
  & Question Answer Rate          & 89.8\% $\pm$ 25.3\% \\
  & Appropriate Response Rate     & 87.0\% $\pm$ 20.6\% \\
\midrule
\multirow{2}{*}{System Efficiency}
  & Error Rate                    & 0.0\% \\
  & Avg.\ Response Latency        & 35.6 s $\pm$ 103.9 s \\
\midrule
\multirow{3}{*}{Device Control ($n$=30)}
  & Information Before Action     & 100.0\% \\
  & Action Explanation Rate       & 94.4\% $\pm$ 17.9\% \\
  & Action Confirmation Rate      & 86.8\% $\pm$ 26.3\% \\
\bottomrule
\end{tabular}\\[0.3em]
{\scriptsize Avg.\ = Average.}
\end{table}
% === END: evaluation_results_table.tex ===

\section{Impact}

% === BEGIN: 04_impact.tex ===
% SECTION 4: IMPACT
% Purpose: Position HEMA in HEM landscape, stakeholder benefits, evaluation contribution
% Target length: 1-1.5 pages

\subsection{HEMA in the Home Energy Management Landscape}

\hema{} occupies a distinct position in the \hems{} landscape as an open-source, multi-agent system
that combines the analytical depth of optimization-driven approaches with the accessibility
of conversational interfaces. By routing queries to specialized agents -- each equipped with
domain-specific tools and reasoning capabilities -- \hema{} delivers targeted, accurate responses
across diverse \hem{} tasks without requiring users to understand the underlying complexity.
This architecture enables a shift from passive energy monitoring toward sustained, informed
human-AI collaboration in residential settings.

\subsection{Stakeholder Benefits}

\textbf{Homeowners \& Consumers}:
\hema{} enables homeowners to engage with their energy data through natural language,
eliminating the need for technical expertise. Users can identify high-consumption appliances,
understand time-of-use rate implications, explore rebate eligibility, and control smart
devices -- all within a single conversational interface. The system adapts its explanations
to user expertise, providing conceptual overviews for newcomers and detailed optimization
strategies for technically proficient users.

\textbf{Utility Companies}:
Utilities benefit from customer-side demand flexibility enabled by \hema{}'s device control
and scheduling capabilities. The system can propose and execute load shifts -- such as
\ev{} charging during off-peak windows or \hvac{} pre-conditioning -- supporting demand
response programs and peak load reduction.

\textbf{Energy Literacy}:
\hema{}'s knowledge agent helps users understand and qualify for utility rebate programs,
increasing program participation. Beyond rebates, the system explains energy concepts
(e.g., TOU pricing, heat pump efficiency, phantom loads) in accessible language,
improving household energy literacy. By enabling users to ask questions and receive
contextualized explanations, \hema{} supports informed decision-making that extends
beyond immediate cost savings toward long-term energy awareness.

\textbf{Researchers \& Academics}:
\hema{} serves as an extensible research platform for the \hem{} community. Its open-source
architecture (GPL-3.0) supports rapid prototyping of new agent specializations, tool
integrations, and reasoning strategies. Researchers can extend the system with custom agents,
integrate alternative \llm{} providers, or modify system prompts to study occupant interaction
patterns and decision-making. 
Furthermore, \hema{}'s built-in evaluation framework -- featuring
an \llm{}-as-simulated-user methodology with 23 objective metrics (Section~2) -- provides
a reusable, reproducible approach for benchmarking \llm{}-integrated \hems{}.
% === END: 04_impact.tex ===

\section{Limitations}

% === BEGIN: 05_limitations.tex ===
% SECTION 5: LIMITATIONS
% Purpose: Discuss current constraints and areas for improvement
% Target length: 0.5-1 page (~250-300 words)

\hema{} has several limitations that should be acknowledged. As an \llm{}-integrated system,
\hema{} inherits the general limitations of \llm{}s, including non-deterministic outputs, 
potential hallucination \cite{Huang2025HallucinationLLM}, and sensitivity to prompt phrasing. 
While the multi-agent architecture with purpose-built tools mitigates these risks by grounding 
responses in structured data and domain-specific logic
, and the Control agent enforces safety guardrails (user confirmation for device control actions),
these measures do not fully eliminate LLM-related risks for safety-critical device control operations. Therefore,
users -- particularly beginners who are 
the target audience -- should be aware that \llm{}-generated explanations and recommendations 
may occasionally be inaccurate or inconsistent across sessions. 
These safety-related features should be thoroughly considered and implemented
when deploying \hema{} in real-world settings.

\hema{} is currently configured as a single-household proof-of-concept using 
static appliance-level energy consumption data (CSV files), 
which may not always be available across different households. 
While the modular architecture supports extension to multi-household deployments by replacing the 
energy data source provided to the \hema{} system, the current version does not include a multi-household configuration
or community-level network management features.

As a system foundation designed for developer extension, the current validation
may not fully reflect real-world device control complexity. As a result,
deploying \hema{} into actual houses would require integration with real-time
data sources (e.g., smart meters, IoT devices, smart electric panels) and utility APIs,
or Model Context Protocols (MCP) for real-time data streaming,
which introduces additional technical complexity and challenges such as 
sensor noise, communication latency, and protocol variability that require future testing and refinement.

Furthermore, reliance on external APIs for \llm{} inference may introduce latency,
vendor dependence, and privacy concerns. Local inference via Ollama, included in \hema{}'s
pipeline, can eliminate these dependencies entirely.

Lastly, the \llm{}-as-simulated-user evaluation methodology may not fully capture
genuine human interaction dynamics, though it enables scalable testing across diverse
personas and scenarios \cite{Yoon2024EvaluatingLLM, Algherairy2025PromptingLLM}.
Human subject studies using \hema{}'s web-based interface are an important direction
for future work.
% === END: 05_limitations.tex ===

\section{Conclusions and Future Work}

% === BEGIN: 06_conclusion.tex ===
% SECTION 6: CONCLUSION AND FUTURE WORK
% Purpose: Summarize contributions and call to action
% Target length: ~200 words

This paper presents \hema{}, the first open-source multi-agent \hems{} that combines
\llm{} reasoning with domain-specific tools to support diverse home energy management tasks
through natural language interaction. Three specialized agents -- Analysis, Knowledge, and Control --
coordinated through a self-consistency classifier, deliver personalized energy insights while
maintaining user control and transparency. Complemented by a reusable evaluation framework with
23 objective metrics and an \llm{}-as-simulated-user methodology, \hema{} provides both a
deployable residential tool and a research platform for advancing \llm{}-integrated \hems{},
addressing a gap in the current literature where evaluation methodologies remain inconsistent
and narrowly scoped. Through practical demonstrations using a real-world dataset of a household's
energy consumption, we show how \hema{} supports informed decision-making and sustained engagement
in \hem{}, highlighting its potential as a user-friendly, adaptable, and effective tool for
residential deployment and \hems{} research.

Future developments include expanding the tool ecosystem, enhancing query
routing with more advanced classification techniques, developing multi-agent coordination
strategies for complex multi-step tasks, and integrating real-time data sources
(e.g., smart meters, IoT devices, utility APIs) to enable dynamic, model predictive control.
Planned work also includes deploying \hema{} in real-world settings with actual occupants to
validate human-AI interaction effectiveness, and extending the system to multi-household and
community-level energy management.
% === END: 06_conclusion.tex ===

\acknowledgments{This material is based upon work partially supported by the National Science Foundation (NSF) under grant 
\#2519054. Any opinions, findings, and conclusions or recommendations expressed in this material are 
those of the author(s) and do not necessarily reflect the views of the NSF.}
\availability{\hema{} is publicly available at: \url{https://github.com/humanbuildingsynergy/HEMA}

% Code Availability: Reproducibility table (R2.4)

Table~\ref{tab:reproducibility} clarifies the reproducibility boundaries of \hema{}.
All software components (frontend, backend, agents, evaluation framework) run locally.
The repository includes sample energy consumption data, device configuration files,
utility rate schedules, evaluation personas and scenarios, and system prompts.
An \llm{} API key (OpenAI, Google, or Anthropic) is required for agent reasoning;
alternatively, local inference via Ollama eliminates external API dependencies entirely.
Users can expect to reproduce the full conversational pipeline, tool outputs, and
evaluation metrics with their own API credentials.

\begin{table}
\caption{{Reproducibility boundaries: local vs.\ external components}}
\label{tab:reproducibility}
\centering
\small
\begin{tabular}{p{4.0cm}cp{5.5cm}}
\toprule
\textbf{Component} & \textbf{External API} & \textbf{Included Data} \\
\midrule
Frontend (React) & No & N/A \\
Backend (FastAPI) & No & N/A \\
Analysis Agent & LLM API & Sample energy and utility rate CSVs \\
Knowledge Agent & LLM API & DOE, ENERGY STAR docs \\
Control Agent & LLM API & Device config JSON \\
Evaluation framework & LLM API & Personas \& scenarios \\
\bottomrule
\end{tabular}
\end{table}}
\aiuse{During the preparation of this work the author used ChatGPT 5.2, and Claude Opus 4.5 and 4.6 in order to 
proofread sentences. After using these tools/services, the author reviewed and edited the content
as needed and takes full responsibility for the content of the publication.}

\appendix
\numberwithin{table}{section}
\numberwithin{figure}{section}

\section{Agent Tools and Capabilities}

% === BEGIN: appendix_a_tools.tex ===
% APPENDIX A: AGENT TOOLS AND CAPABILITIES

This appendix provides a detailed inventory of the purpose-built tools available to each
of \hema{}'s three specialized agents (described in Section 2). Each tool is implemented
as a Python function that the agent can invoke during the \react{} reasoning cycle. The
agent autonomously selects which tools to call, determines appropriate parameters, and
interprets the results to formulate its response. Developers can extend the system by
adding new tools following the same interface pattern.

\subsection{Analysis Agent Tools}

The Analysis agent includes 18 tools organized into four categories as shown in
Table~\ref{tab:analysis_tools}. Data loading tools handle ingestion of appliance-level
energy consumption data. Consumption and period analysis tools perform statistical
computations and comparisons across flexible time ranges. Pattern and trend analysis
tools detect usage patterns, variability, and temporal trends. Rate, solar, and summary
tools integrate utility rate structures and renewable energy alignment to quantify cost
impacts and identify savings opportunities.

\begin{table}
\caption{Analysis Agent tools and descriptions}
\label{tab:analysis_tools}
\centering
\small
\begin{tabular}{p{5.0cm}p{9.5cm}}
\toprule
\textbf{Tool Name} & \textbf{Description} \\
\midrule
\multicolumn{2}{c}{\textit{Data Loading}} \\
\midrule
load\_energy\_data & Load appliance-level energy consumption CSV file \\
list\_available\_data & List all available energy data files and rate files \\
get\_tracked\_appliances & List all appliances being monitored in dataset \\
\midrule
\multicolumn{2}{c}{\textit{Consumption and Period Analysis}} \\
\midrule
analyze\_consumption & Overall consumption patterns (daily/hourly profiles, peak times, statistics) \\
analyze\_appliances & Appliance-level breakdown; rank by energy usage \\
query\_energy\_data & Query energy data with flexible time ranges and filters \\
compare\_energy\_periods & Compare consumption between two time periods; calculate differences \\
analyze\_energy\_period & Analyze consumption for any time period with flexible aggregation \\
\midrule
\multicolumn{2}{c}{\textit{Pattern and Trend Analysis}} \\
\midrule
calculate\_rolling\_average & Calculate rolling average consumption with trend analysis \\
compare\_weekday\_weekend & Compare weekday vs.\ weekend energy consumption patterns \\
analyze\_peak\_hours & Analyze peak vs.\ off-peak consumption with savings estimates \\
analyze\_usage\_frequency & Identify appliance usage frequency patterns \\
analyze\_usage\_variability & Analyze appliance usage variability using coefficient of variation \\
\midrule
\multicolumn{2}{c}{\textit{Rate, Solar, and Summary}} \\
\midrule
get\_utility\_rate & Retrieve utility rate structure (flat or time-of-use) \\
analyze\_utility\_rate & Analyze energy usage relative to rate structure (TOU or flat) \\
analyze\_solar\_availability & Analyze solar power generation patterns and availability \\
analyze\_solar\_alignment & Measure appliance usage alignment with solar generation \\
get\_analysis\_summary & Generate comprehensive summary of all analysis results \\
\bottomrule
\end{tabular}
\end{table}

\subsection{Knowledge Agent Tools}

The Knowledge agent includes 8 tools for information retrieval and domain knowledge as
shown in Table~\ref{tab:knowledge_tools}. Retrieval-augmented generation tools enable
semantic search over indexed energy documents (U.S. Department of Energy guides, ENERGY STAR specifications,
utility rebate databases) to provide grounded, citation-aware responses. Weather
integration tools retrieve real-time and forecast weather data to support context-aware
energy consumption guidance. The User Context tool enables personalized recommendations
based on the user's loaded energy data.

\begin{table}
\caption{Knowledge Agent tools and descriptions}
\label{tab:knowledge_tools}
\centering
\small
\begin{tabular}{p{5.0cm}p{9.5cm}}
\toprule
\textbf{Tool Name} & \textbf{Description} \\
\midrule
\multicolumn{2}{c}{\textit{Retrieval-Augmented Generation}} \\
\midrule
energy\_knowledge & Explain energy concepts, best practices, tips (TOU, heat pumps, solar, etc.) \\
search\_energy\_documents & Search indexed documents (DOE guides, ENERGY STAR, rebate info) \\
get\_knowledge\_base\_status & Check RAG knowledge base status and available sources \\
\midrule
\multicolumn{2}{c}{\textit{Weather Integration}} \\
\midrule
get\_current\_weather & Retrieve current weather conditions (temperature, humidity, wind, cloud) \\
get\_weather\_forecast & Provide 1--7 day forecast with temperature, conditions, precipitation \\
get\_weather\_energy\_impact & Analyze how current and forecasted weather affects energy usage \\
get\_historical\_weather & Retrieve historical weather data for past date ranges \\
\midrule
\multicolumn{2}{c}{\textit{User Context}} \\
\midrule
get\_user\_context & Get current user's energy data context for personalized advice \\
\bottomrule
\end{tabular}
\end{table}

\subsection{Control Agent Tools}

The Control agent includes 10 tools for device management and automation as shown in
Table~\ref{tab:control_tools}. Device discovery and status tools enable the agent to
inventory available smart devices and query their current state before taking action.
Control actions tools execute commands such as adjusting thermostat setpoints, changing
operating modes, or toggling power states -- with parameter validation against device
operational limits. Scheduling and automation tools support time-based device programming
for load shifting and demand response. Two shared utility tools (also used by the
Analysis and Knowledge agents, respectively) provide rate structure and weather context
for energy-aware device optimization.

\begin{table}
\caption{Control Agent tools and descriptions}
\label{tab:control_tools}
\centering
\small
\begin{tabular}{p{5.0cm}p{9.5cm}}
\toprule
\textbf{Tool Name} & \textbf{Description} \\
\midrule
\multicolumn{2}{c}{\textit{Device Discovery and Status}} \\
\midrule
get\_device\_list & List all available smart devices with status summary \\
get\_device\_status & Get detailed status of specific device (temperature, mode, power state) \\
get\_all\_devices\_energy & Get energy consumption across all devices \\
\midrule
\multicolumn{2}{c}{\textit{Control Actions}} \\
\midrule
control\_device & Execute control commands (adjust setpoints, change modes, power on/off) \\
get\_available\_actions & List available control actions for specific device \\
get\_device\_energy & Get power consumption for specific device \\
\midrule
\multicolumn{2}{c}{\textit{Scheduling and Automation}} \\
\midrule
schedule\_device\_action & Schedule device action for specific time (e.g., charge at midnight) \\
get\_automation\_rules & List configured automation rules and triggers \\
\midrule
\multicolumn{2}{c}{\textit{Shared Utility}} \\
\midrule
get\_utility\_rate & Retrieve utility rate structure for TOU-aware scheduling \\
get\_current\_weather & Retrieve current weather for HVAC optimization decisions \\
\bottomrule
\end{tabular}
\end{table}
% === END: appendix_a_tools.tex ===

\section{Agent System Prompts}

% === BEGIN: appendix_b_prompts.tex ===
% APPENDIX B: AGENT SYSTEM PROMPTS

This appendix provides excerpts of the system prompts that define each agent's role,
capabilities, workflow, and constraints. These prompts are central to \hema{}'s behavior,
as they instruct the \llm{} on how to reason, which tools to invoke, and what guardrails
to respect. Full prompts are stored in the \texttt{prompts/}
directory of the repository and can be customized by developers to modify agent behavior
or adapt the system to specific household configurations.

\paragraph{Analysis Agent System Prompt}
The Analysis Agent prompt specifies workflow rules (data loading requirements, tool
selection guidelines, parameter conventions), adaptive communication strategies
(inferring user technical level and adjusting response style accordingly), and
response guidelines that enforce accuracy and prohibit data fabrication.

\begin{figure}
\centering
\includegraphics[width=0.82\textwidth]{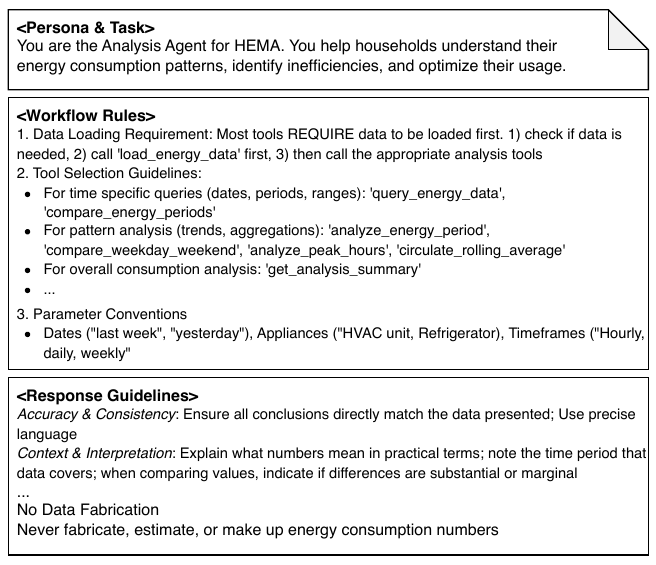}
\caption{\hema{} Analysis Agent system prompt. Some parts were truncated for brevity. 
The full prompt is available in the repository.}
\label{fig:analysis_agent_prompt}
\end{figure}

\paragraph{Knowledge Agent System Prompt}
The Knowledge Agent prompt directs the agent to ground explanations in authoritative
sources (DOE, ENERGY STAR) and integrate weather data for context-aware guidance,
while respecting scope boundaries by referring device control queries to the Control Agent.

\begin{figure}
\centering
\includegraphics[width=0.82\textwidth]{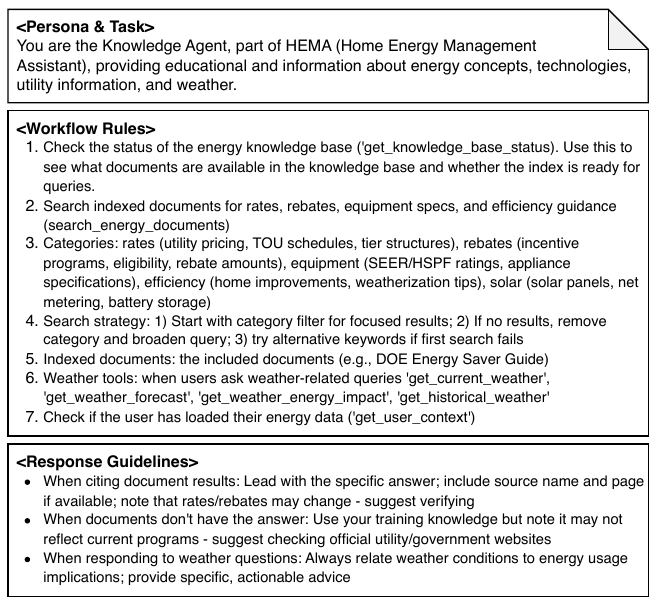}
\caption{\hema{} Knowledge Agent system prompt. Some parts were truncated for brevity. 
The full prompt is available in the repository.}
\label{fig:knowledge_agent_prompt}
\end{figure}

\paragraph{Control Agent System Prompt}
The Control Agent prompt enforces a discovery-first workflow (discover, validate,
execute, confirm) with safety-critical constraints including device operational
limits and mandatory user confirmation before executing significant changes.

\begin{figure}
\centering
\includegraphics[width=0.82\textwidth]{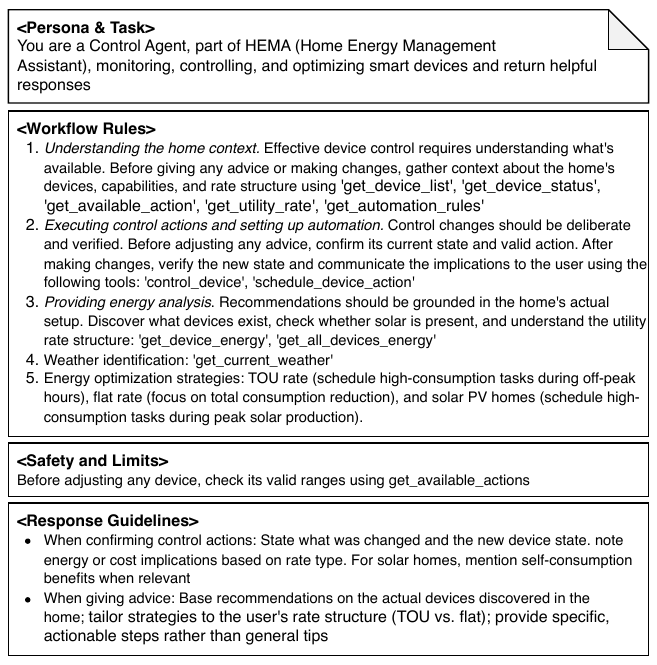}
\caption{\hema{} Control Agent system prompt. Some parts were truncated for brevity. 
The full prompt is available in the repository.}
\label{fig:control_agent_prompt}
\end{figure}
% === END: appendix_b_prompts.tex ===

\section{Evaluation Methodology: Personas and Scenarios}

% === BEGIN: appendix_c_evaluation.tex ===
% APPENDIX B: EVALUATION METHODOLOGY: PERSONAS AND SCENARIOS
% LLM-as-Simulated-User Evaluation Framework

This appendix details the user personas and evaluation scenarios used in \hema{}'s
evaluation framework (described in Section 2). The evaluation employs an \llm{}-as-simulated-user
methodology, where an \llm{} assumes defined persona characteristics to interact with
\hema{} in multi-turn conversations, enabling systematic and reproducible testing without
requiring extensive human subject recruitment.

\paragraph{User Personas}

Three distinct personas were designed to represent different user archetypes commonly
encountered in residential energy management, spanning a range of technical expertise
from novice to expert as shown in Table~\ref{tab:personas}. Each persona is defined by
demographics, background knowledge, communication style, typical behaviors, and
operational constraints. These profiles guide the simulated user's language, question
complexity, and follow-up patterns during evaluation conversations, ensuring that
\hema{}'s ability to adapt responses to different expertise levels is systematically tested.

\begin{table}
\caption{Evaluation personas: three user archetypes with distinct profiles and interaction patterns}
\label{tab:personas}
\centering
\small
\begin{tabular}{p{2.5cm}p{2.5cm}p{3.5cm}p{5.0cm}}
\toprule
\textbf{Persona} & \textbf{Technical Level} & \textbf{Background} & \textbf{Key Characteristics} \\
\midrule
\textbf{Confused Newcomer} & Novice & First-time homeowner (3 months), received high utility bill, 
unfamiliar with energy concepts & Asks clarifying questions, prefers concrete examples, expresses 
uncertainty, limited time available \\
\midrule
\textbf{Tech-Savvy Optimizer} & Expert & Software engineer, loves data analysis, recently installed 
smart thermostat, interested in automation & Requests detailed numbers, asks for comparisons, 
challenges oversimplified responses, wants to understand ``why'' \\
\midrule
\textbf{Budget-Conscious Parent} & Intermediate & Single parent with 2 kids, looking for practical 
cost-reduction solutions, needs time-efficient approaches & Frequently asks ``how much will 
I save?'', focuses on practical solutions, compares options by cost-benefit \\
\bottomrule
\end{tabular}
\end{table}

\paragraph{Evaluation Scenarios}

Seven core scenarios were developed to comprehensively test \hema{}'s capabilities across
all three agent types as shown in Table~\ref{tab:scenarios}. Each scenario defines a realistic
\hem{} task with a primary user goal and measurable success criteria. The scenarios were
selected to cover representative tasks that households commonly face: understanding energy
bills and rate structures, identifying high-consumption appliances, optimizing device
schedules, and learning about available rebates and energy-saving technologies. Together,
they exercise the full range of \hema{}'s multi-agent capabilities, from data-driven
analysis and device control to educational support.

\begin{table}
\caption{Evaluation scenarios across three agent types}
\label{tab:scenarios}
\centering
\small
\begin{tabular}{p{3.5cm}p{5.5cm}p{5.5cm}}
\toprule
\textbf{Scenario} & \textbf{Primary Goal} & \textbf{Success Criteria} \\
\midrule
\multicolumn{3}{c}{\textit{Analysis Agent (4 scenarios)}} \\
\midrule
\textbf{Understanding My Utility Rate} & Understand \tou{} pricing structure and its impact 
on personal bill & Understand pricing in simple terms, know peak vs. off-peak hours, get 
actionable money-saving tip \\
\midrule
\textbf{Appliance Energy Analysis} & Identify which appliances consume the most energy & 
Rank top energy consumers, understand relative usage differences, get upgrade recommendations \\
\midrule
\textbf{Peak Usage Reduction} & Develop strategy to shift energy usage away from expensive 
peak hours & Understand peak contributors, receive concrete implementation plan, learn potential savings \\
\midrule
\textbf{Multi-Step Investigation} & Investigate complex question requiring multiple 
analysis angles (why costs vary across days) & Receive multi-perspective analysis, 
understand root causes, get recommendations based on findings \\
\midrule
\multicolumn{3}{c}{\textit{Control Agent (2 scenarios)}} \\
\midrule
\textbf{Thermostat Adjustment} & Adjust smart thermostat to optimal settings based on rate 
structure for energy savings & Successfully change temperature, understand impact on costs, 
get confirmation of changes \\
\midrule
\textbf{Multi-Device Vacation Prep} & Configure multiple devices (HVAC, water heater, pool 
pump) for minimal energy use during 2-week vacation & Configure HVAC for minimal usage, 
set water heater to vacation mode, confirm all changes \\
\midrule
\multicolumn{3}{c}{\textit{Knowledge Agent (1 scenario)}} \\
\midrule
\textbf{Energy Rebate Information} & Learn about available rebates and incentive programs 
for energy-efficient upgrades & Learn specific rebate programs, understand eligibility 
requirements, know rebate amounts and application process \\
\bottomrule
\end{tabular}
\end{table}

\subsection{Evaluation Coverage}

The combination of 3 personas and 7 scenarios creates 21 base scenario combinations
(3 $\times$ 7), ensuring that each scenario is tested across all expertise levels. Each
combination is executed 5 times to capture the inherent variability of \llm{}-generated
responses and compute standard deviations, yielding 105 total evaluation runs as shown
in Table~\ref{tab:evaluation_coverage}. This structured approach ensures comprehensive
coverage of \hema{}'s three specialized agents across diverse user profiles and realistic
\hem{} tasks.

\begin{table}
\caption{Evaluation coverage: personas, scenarios, and repetitions}
\label{tab:evaluation_coverage}
\centering
\small
\begin{tabular}{p{9.0cm}c}
\toprule
\textbf{Evaluation Component} & \textbf{Count} \\
\midrule
User Personas & 3 \\
Evaluation Scenarios & 7 \\
Analysis Agent Scenarios & 4 \\
Control Agent Scenarios & 2 \\
Knowledge Agent Scenarios & 1 \\
Base Scenario Combinations (Persona $\times$ Scenario) & 21 \\
Repetitions per Combination & 5 \\
Total Evaluation Runs & 105 \\
\bottomrule
\end{tabular}
\end{table}
% === END: appendix_c_evaluation.tex ===

\section{{Evaluation Metrics}}

% === BEGIN: appendix_d_metrics.tex ===
% Appendix D: Evaluation Metrics (R2.1)

This appendix defines the 23 objective evaluation metrics used to assess \hema{}.
All metrics are automatically computed without subjective LLM judgment.

Table~\ref{tab:metric_defs} organizes the metrics by category.
Tier~1 metrics use pure counting; Tier~2 metrics use LLM-based extraction with
deterministic scoring; Tier~3 metrics verify numerical claims against ground truth data.

\begin{table}
\caption{Definitions of the 23 objective evaluation metrics.}
\label{tab:metric_defs}
\centering
\scriptsize
\begin{tabular}{rp{4.2cm}p{7.8cm}}
\toprule
\textbf{\#} & \textbf{Metric} & \textbf{Description} \\
\midrule
\multicolumn{3}{l}{\textit{Task Performance}} \\
1  & Goal Achievement        & Whether the conversation achieved the stated goal (binary) \\
2  & Turns to Completion     & Number of conversation turns to achieve the goal \\
3  & Task Efficiency         & Ratio of max allowed turns to actual turns used \\
\midrule
\multicolumn{3}{l}{\textit{Factual Accuracy}} \\
4  & Claim Accuracy          & \% of numerical claims within 5\% of ground truth value \\
5  & Mean Error              & Average percentage error across all verified claims \\
6  & Num.\ Factual Claims    & Count of verifiable numerical claims extracted \\
\midrule
\multicolumn{3}{l}{\textit{Interaction Quality}} \\
7  & Question Answer Rate    & \% of user questions that received a direct answer \\
8  & Appropriate Response Rate & \% of responses matching question type (data-specific vs.\ general) \\
9  & Actionable Rec.\ Ratio  & \% of recommendations that are specific and actionable \\
10 & Jargon Explanation Rate  & \% of technical terms that were explained to the user \\
11 & Avg.\ Response Length    & Mean character count of system responses \\
12 & Data Sources Referenced  & Count of data sources cited in responses \\
13 & User Questions Count     & Total questions asked by the user \\
\midrule
\multicolumn{3}{l}{\textit{System Efficiency}} \\
14 & Avg.\ Response Latency   & Mean response time per system turn (s) \\
15 & Error Rate               & \% of turns that produced errors \\
16 & Tool Calls per Run       & Total tool invocations during conversation \\
17 & Tokens per Run           & Total input + output tokens consumed \\
18 & Cost per Run             & Total API cost in USD \\
\midrule
\multicolumn{3}{l}{\textit{Device Control}} \\
19 & Info.\ Before Action     & \% of control actions preceded by information gathering \\
20 & Action Confirmation Rate & \% of control actions confirmed with the user \\
21 & Action Explanation Rate  & \% of control actions accompanied by an explanation \\
22 & Control Tools Called     & Count of device control tool invocations \\
23 & Info.\ Tools Called      & Count of information-gathering tool invocations \\
\bottomrule
\end{tabular}\\[0.3em]
{\scriptsize Num.\ = Number; Avg.\ = Average; Rec.\ = Recommendation; Info.\ = Information.}
\end{table}

% === END: appendix_d_metrics.tex ===

\section{{Detailed Evaluation Results}}

% === BEGIN: appendix_e_results.tex ===
% Appendix E: Detailed Evaluation Results (R2.1)

This appendix presents detailed results from 105 test runs
(3 personas $\times$ 7 scenarios $\times$ 5 repetitions per combination).
The seven scenarios cover four Analysis, two Control, and one Knowledge task;
future work will expand Knowledge Agent evaluation across additional scenarios.

\subsection{Per-Scenario Results}

Table~\ref{tab:per_scenario} breaks down key metrics by scenario.
All scenarios achieved 100\% goal achievement except vacation preparation (86.7\%),
where 2 of 15 runs reached the maximum turn limit before completing all sub-goals.
Factual accuracy was highest for knowledge-oriented scenarios
(rebate inquiry and utility rate: 100\%) and lowest for peak reduction strategy (91.1\%),
where some runs produced approximate calculations that exceeded the 5\% error tolerance.
The two failures in vacation preparation resulted from the scenario's
multi-device complexity -- requiring sequential HVAC and water heater
reconfiguration with user confirmation -- which occasionally exceeded
the turn budget when simulated users requested additional clarifications.
The high variance in turns to completion (6.3 $\pm$ 9.5) reflects the
range of scenario complexity, from simple 2-turn rebate inquiries to
14-turn multi-device orchestration.
Average response latency (35.6\,s $\pm$ 103.9\,s) also varied substantially:
device control and energy analysis scenarios exhibited moderate latency (14--22\,s),
while multi-device orchestration (e.g., vacation preparation) required
longer processing (118.3\,s) due to sequential tool invocations.

\smallertable
\begin{table}\setlength{\tabcolsep}{5.5pt}
\caption{Per-scenario evaluation results (each scenario: $n$=15, i.e., 3 personas $\times$ 5 runs).}
\label{tab:per_scenario}
\centering
\scriptsize
\begin{tabular}{@{}lrrrrrrr@{}}
\toprule
 & \rotatebox{60}{\parbox{1.75cm}{\textbf{Appliance\\analysis}}} & \rotatebox{60}{\parbox{1.75cm}{\textbf{Multi-step\\analysis}}} & \rotatebox{60}{\parbox{1.75cm}{\textbf{Peak\\reduction}}} & \rotatebox{60}{\parbox{1.75cm}{\textbf{Rebate\\inquiry}}} & \rotatebox{60}{\parbox{1.75cm}{\textbf{Thermostat\\adjustment}}} & \rotatebox{60}{\parbox{1.75cm}{\textbf{Utility\\rate}}} & \rotatebox{60}{\parbox{1.75cm}{\textbf{Vacation\\preparation}}} \\
\midrule
Goal (\%)              & 100.0 & 100.0 & 100.0 & 100.0 & 100.0 & 100.0 &  86.7 \\
Turns (mean)           &   3.9 &   4.3 &  11.3 &   2.1 &   6.7 &   3.3 &  13.6 \\
Question answer (\%)   &  82.8 &  86.7 &  91.3 & 100.0 &  89.5 &  85.3 &  93.3 \\
Approp.\ response (\%) &  76.0 &  90.7 &  93.1 & 100.0 &  91.9 &  75.5 &  82.2 \\
Factual acc.\ (\%)     &  96.8 &  94.1 &  91.1 & 100.0 &  99.1 & 100.0 &  96.3 \\
Latency (s)            &  22.2 &  48.1 &  19.4 &  11.6 &  14.8 &  14.6 & 118.3 \\
\bottomrule
\end{tabular}\\[0.3em]
{\scriptsize Approp.\ = Appropriate; acc.\ = accuracy.}
\end{table}

\subsection{Per-Persona Results}

Table~\ref{tab:per_persona} summarizes results by user persona.
All three personas achieved comparable performance, with the budget-conscious parent
reaching 100\% goal achievement. The tech-savvy optimizer exhibited higher
appropriate response rates (91.3\%), reflecting \hema{}'s ability to adapt its
communication style to the user's technical expertise.

\begin{table}
\caption{Per-persona evaluation results (each persona: $n$=35, i.e., 7 scenarios $\times$ 5 runs).}
\label{tab:per_persona}
\centering
\small
\begin{tabular}{lrrrrr}
\toprule
\textbf{Persona} & \textbf{Goal} & \textbf{Turns} & \textbf{QA} & \textbf{Approp.} & \textbf{Latency} \\
                  & \textbf{(\%)} &                & \textbf{(\%)} & \textbf{(\%)} & \textbf{(s)} \\
\midrule
Budget-conscious parent  & 100.0 &  4.9 & 88.0 & 85.4 & 20.7 \\
Confused newcomer        &  97.1 &  8.7 & 94.3 & 84.5 & 22.2 \\
Tech-savvy optimizer     &  97.1 &  5.4 & 87.3 & 91.3 & 63.9 \\
\bottomrule
\end{tabular}\\[0.3em]
{\scriptsize QA = Question Answer Rate; Approp.\ = Appropriate Response Rate.}
\end{table}

\subsection{Interaction Quality and System Efficiency}

Table~\ref{tab:quality_detail} presents the full set of interaction quality metrics.
The actionable recommendation ratio (64.1\%) reflects \hema{}'s design to balance
specific, data-backed recommendations with educational context and general
energy management guidance -- particularly for knowledge-oriented queries.
The jargon explanation rate (44.2\% $\pm$ 43.5\%) exhibits high variance
because it is strongly scenario-dependent: knowledge-oriented queries
involving technical concepts (e.g., \tou{} pricing) yield high explanation
rates, while device control scenarios use minimal technical terminology,
producing near-zero rates and a bimodal distribution across runs.

\begin{table}
\caption{Detailed interaction quality and system efficiency metrics ($n$=105).}
\label{tab:quality_detail}
\centering
\small
\begin{tabular}{llr}
\toprule
\textbf{Category} & \textbf{Metric} & \textbf{Mean $\pm$ Std.\ Dev.} \\
\midrule
\multirow{4}{*}{Interaction Quality}
  & Question Answer Rate            & 89.8\% $\pm$ 25.3\% \\
  & Appropriate Response Rate       & 87.0\% $\pm$ 20.6\% \\
  & Actionable Recommendation Ratio & 64.1\% $\pm$ 29.7\% \\
  & Jargon Explanation Rate         & 44.2\% $\pm$ 43.5\% \\
\midrule
\multirow{4}{*}{System Efficiency}
  & Avg.\ Response Latency          & 35.6 s $\pm$ 103.9 s \\
  & Tool Calls per Run              & 18.4 $\pm$ 36.5 \\
  & Tokens per Run                  & 48{,}766 $\pm$ 35{,}036 \\
  & Cost per Run (USD)              & \$0.009 $\pm$ \$0.006 \\
\midrule
\multirow{3}{*}{Device Control ($n$=30)}
  & Information Before Action       & 100.0\% $\pm$ 0.0\% \\
  & Action Explanation Rate         & 94.4\% $\pm$ 17.9\% \\
  & Action Confirmation Rate        & 86.8\% $\pm$ 26.3\% \\
\bottomrule
\end{tabular}
\end{table}

% === END: appendix_e_results.tex ===

\bibliographystyle{unsrtnat}
\bibliography{references}

\end{document}